# LANC: locality-aware network coding for better P2P traffic localization

Guoqiang Zhang*, Suqi Cheng*[†] and Guoqing Zhang*
*Institute of Computing Technology, Chinese
Academy of Sciences, Beijing, China.
[†]Graduate University of Chinese Academy of Sciences, Beijing, China.
Email:{guoqiang, chengsuqi, gqzhang}@ict.ac.cn

*Abstract*—As ISPs begin to cooperate to expose their network locality information as services, e.g., P4P, solutions based on locality information provision for P2P traffic localization will soon approach their capability limits. A natural question is: can we do any better provided that no further locality information improvement can be made? This paper shows how the utility of locality information could be limited by conventional P2P data scheduling algorithms, even as sophisticated as the local rarest first policy.

Network coding's simplified data scheduling makes it competent for improving P2P application's throughput. Instead of only using locality information in the topology construction, this paper proposes the locality-aware network coding (LANC) that uses locality information in both the topology construction and downloading decision, and demonstrates its exceptional ability for P2P traffic localization. The randomization introduced by network coding enhances the chance for a peer to find innovative blocks in its neighborhood. Aided by proper locality-awareness, the probability for a peer to get innovative blocks from its proximity will increase as well, resulting in more efficient use of network resources. Extensive simulation results show that LANC can significantly reduce P2P traffic redundancy without sacrificing application-level performance. Aided by the same locality knowledge, the traffic redundancies of LANC in most cases are less than 50% of the current best approach that does not use network coding.

*Keywords:*locality-aware network coding, P2P traffic localization, network coding, locality-awareness, beyond P4P

## I. INTRODUCTION

In recent years, peer-to-peer content distribution applications, both bulk data file distribution and real-time video streaming, e.g., BitTorrent [1] and PPLive [2], have emerged as promising solutions and attracted wide-spread attentions among Internet users due to its scalability with user population and resilience to flash crowds as well as dynamic user arrivals and departures. However, they also pose serious challenges to the network operators because of the heavy traffic generated by them. Measurement data show that P2P applications have already become the dominant network bandwidth consumer, contributing more than 50% to the overall network traffic [3], [4]. In some cases, the P2P traffic consumes so much bandwidth that it starves other services [5].

One major cause of the tremendous traffic is the inefficient use of network resources. P2P systems are based on an overlay topology on top of the underlying network, upon which content searching, and sometimes routing, is performed. This leads to problems for both the P2P applications and network operators. Being largely network-oblivious, overlay network construction either disregards or relies on limited knowledge of underlying network topology, which inevitably leads to inefficient overlay topology and routing, and consequently sub-optimal application level performance. The problems for network operators are twofold: (1) the dynamics of P2P traffic makes ISPs hard to manage these traffic and unable to do effective traffic engineering [6], [7]; and (2) P2P traffic often crosses network boundaries multiple times [8], [9], which can either incur increased costs for its inter-domain traffic or cause traffic imbalance with its peers, potentially violating the contractual peering agreements [7]. In addition to the possibly rising cost and contract violation, the increase of inter-domain traffic is even more serious in that inter-domain links, rather than the ISP backbone links, are the major source of network bottlenecks [10].

ISPs' tremendous investment in upgrading bandwidth proves to be effortless in face of the endless and relentless traffic demands of the P2P applications. The solutions towards relieving the P2P traffic burden can be coarsely categorized into three phases. In the initial phase, the tough reality makes some ISPs to unilaterally throttle or rate limit P2P applications using techniques such as deep packet inspection(DPI). In reaction, the P2P application developers use techniques such as dynamic ports or message encryption to hide their traffic. Hence, a vicious circle emerges, which is dramatized as a *war* between the P2P content providers and ISPs in the IETF. In the second phase, P2P application developers and ISPs begin to independently seek approaches to localize the P2P traffic. The P2P applications can obtain coarse-grained locality knowledge via reverse-engineering to aid its overlay network building and downloading [9], [13]–[18], whereas the ISPs can leverage the widely deployed caches to localize the P2P traffic [9], [11], [12]. Recently and in the third phase, P2P content providers and ISPs begin to cooperate to make efficient use of the network resources [6], [7], which is believed to be the only way to relieve the tension between them, e.g., P4P.

As ISPs begin to provide locality information as network services, the performance gain arising from the accuracy of network locality will soon approach its limit. A natural question arises: can we do any better, i.e., more efficient network resource utilization, beyond P4P? The chance lies in the way how data is scheduled and propagated in the network. Traditional P2P paradigm either uses random policy or relies on local information, e.g., local rarest first policy, to make data scheduling decisions, which may lead to biased distribution of data blocks in the network. As a result, no matter how accurate the locality information is, the data scheduling and propagation mode limits further performance improvement.

Network coding [24] allows intermediate nodes not only to receive, replicate and forward, but also to do arbitrary coding operations on packets. By applying network coding to P2P content distribution, the data blocks transmitted on the network are no longer the original blocks, but coded blocks of multiple random original blocks. This way, network coding greatly simplifies the data scheduling and enhances the application-level throughput [32], [34]. However, it is argued that the computation overhead incurred by the encoding and decoding processes may offset or even outweigh the benefits [25], which significantly impedes its wide adoption and deployment. In addition, these studies are all carried out from the perspective of improving application-level performance, without consideration for ISPs' interest.

However, standing at ISPs' point of view, the randomization introduced by network coding improves the probability for a peer to find innovative blocks from its neighborhood, which, if aided by proper locality-aware overlay construction and downloading, can potentially increase the chance of finding and retrieving innovative blocks from its proximity. Therefore, network coding could be a powerful tool to enhance the efficiency of network resource utilization, with the ability to breach the capability limits of state-of-art techniques. The main contributions of this paper are listed as follows:

1) We show how the utility of locality information could be limited by conventional data scheduling algorithms, whereas this limitation can be greatly avoided by network coding based scheduling;
2) We propose a pull-based network coding model and show how integrating the locality-awareness and network coding(LANC) could significantly reduce P2P traffic redundancy. Simulation results show that aided by the same locality knowledge, the traffic redundancy of LANC in most cases can be reduced to less than 50% of the current best approach;
3) We provide considerations for a practical LANC implementation, such as optimal parameter settings and balancing between the computation overhead and the system throughput ;

The subsequent paper is organized as follows: Related work of network coding and P2P traffic localization are discussed in section II. A simple motivation example is presented in section III to illustrate how data scheduling affects and how network coding improves P2P traffic localization. We present the model used for our study in section IV and the experimental evaluation results in section V. Implementation considerations are discussed in section VI. Finally, we conclude the paper in section VII.

## II. Related Work

### A. network coding

Network coding is a paradigm shift from the conventional information transmission and processing mode. By allowing the intermediate nodes to perform arbitrary coding function on the input data, it emerges as a promising technology to realize the theoretical multicast capacity upper bound predicted by the max-flow min-cut theorem [24]. It is demonstrated that linear codes on Galois finite field can achieve this rate [26], [27].

The concept of random network coding [29] paves the way for the practicality of network coding. Random network coding allows each node to autonomously select its code, as opposed to the centralized assignment in deterministic algorithms [26], [28], hence facilitating the distributed implementation in a large and dynamic environment. Consequently, there is a gradual research shift from the theoretical point of view to the more practical settings [30]–[35]. Avalanche has demonstrated that network coding can significantly improve the throughput and robustness of the system due to its simplified data scheduling [32], [35]. Lava provides a reality check for the possible performance gain of network coding over P2P real-time video streaming [34].

On the other hand, it is argued that the computation overhead incurred by encoding and decoding cannot be neglected so that the use of network coding should be cautious [25]. In addition, due to the simplified scheduling algorithm, it is believed that the push-based data propagation mode can take full advantage of network coding for throughput improvement, however, even random network coding can cause linearly dependent redundant blocks due to the small-world network topology [36].

### B. P2P traffic localization

Early day's P2P applications are often network-oblivious, with the only aim to optimize the application-level performance, without any consideration of the underlying network operator's interest.

As the P2P traffic quickly becomes the dominant consumer of ISPs' network bandwidth, conflicts arises and changes inevitably emerges, though passively. In this phase, P2P applications and ISPs both unilaterally seek to address the inefficient use of network resources. P2P applications have developed various reverse-engineering techniques to build locality-aware overlay networks and enable locality-aware downloading. These reverse-engineering techniques can be either based on active probing, e.g.,landmark-based proximity identification [16], [17] and network coordinate systems [21]–[23], or on passive inference, e.g., identifying a host's ASN by its IP address [9], [14]. Reverse engineering techniques, however, are inherently limited by the granularity and accuracy of these information.

The ISPs, on the other hand, can rely on their widely deployed caches [9], [11], [12] to cache P2P traffic so that the duplicated transmissions on backbone networks can be reduced. P2P's distinguished traffic characteristic from Web traffic triggers the development of different cache replacement algorithms, e.g., partial caching [11]. However, cache solution is not scalable as it has to speak the appropriate protocols, and it violates the economic rule since ISPs may regard it as a shift of distribution cost from the P2P content providers to themselves [9]. In addition, caching content may raise the legal liability issue.

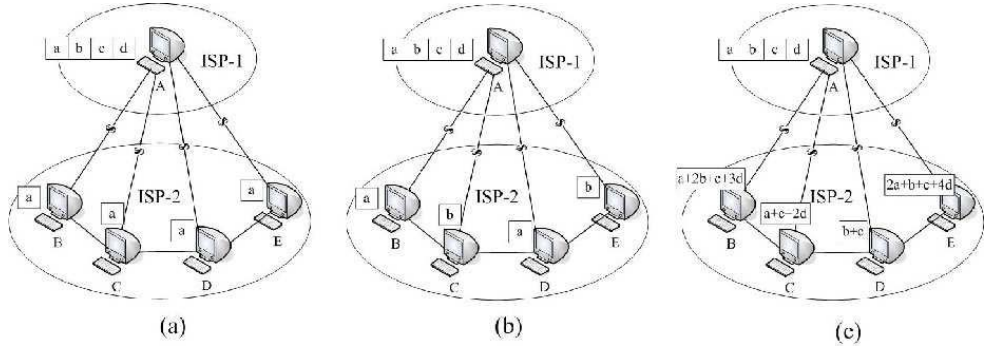

Fig. 1: A motivation example showing how data scheduling algorithms affect the P2P traffic localization. In this figure, (a) data scheduling with random policy, (b) data scheduling with local rarest first policy, and (c) data scheduling with network coding.

Gradually, it becomes a consensus that cooperation between these two sides is the only way to ensure a sustainable development because ISPs are at the right position to offer the most accurate locality knowledge. The Oracle service provides a peer ranking service based on certain topological metrics [6]. P4P proposes an architecture for the ISPs to opaquely expose network distance information without sacrificing its privacy. These distances can then be used by P2P applications to shape their connectivity and choose network-efficient communication patterns [7]. These work provide the means for ISPs and P2P applications to jointly optimize their respective performances. With ISPs' participation, it is expected the locality accuracy and subsequently the performance gain arising from locality information will approach its limit.

To summarize, most previous efforts for P2P traffic localization can be viewed from two orthogonal dimensions, i.e., caching the P2P traffic and enhancing the locality-awareness, except for [14], in which the impact of local rarest first data scheduling policy on P2P traffic localization is investigated. It is shown that by integrating locality knowledge and local rarest first data scheduling, the inter-domain P2P traffic redundancy can be significantly reduced. The data scheduling algorithm is indeed another dimension that could have significant impact on the P2P traffic localization. In [32], it was shown that network coding performs better than not using coding in clustered topologies, however, the locality information is only used for overlay topology construction, but not used in the downloading decision phase. In addition, this work focuses on the application level performance, and doesn't present a quantitative evaluation of the traffic redundancy.

## III. A MOTIVATION EXAMPLE

To show how the data scheduling algorithm will influence the P2P traffic redundancy and how network coding can effectively reduce the traffic redundancy, we use a simple motivation example for illustration purpose, see Fig. 1. In this illustrative figure, $A$-$E$ are five application-layer peers. $A$ is located in ISP-1 and $B$-$E$ are located in ISP-2. The neighboring relationship is given as shown in the figure. Suppose that in the initial stage, $A$ has four blocks $a,b,c,d$, and other peers have no available blocks. Fig .1(c) uses network coding, while the former two do not use network coding.

In Fig .1(a), blocks are requested in a random way, so in the worst case, $B,C,D,E$ could request the same block, say $a$, from peer $A$. In this worst case, the intra-domain links cannot be utilized in the following data request round and all traffic has to pass through the inter-domain links. Certainly, it is possible that $B$-$E$ may request distinct blocks from $A$ so that only one copy of the original blocks has to pass through the inter-domain links and peers in ISP-2 can then exchange blocks between themselves to assemble the four blocks by locality-aware downloading. However, the probability is very low, only $4!/4^4 = 9.375\%$ with this random policy.

In Fig. 1(b), a more sophisticated policy, the local rarest first policy, is used. However, even this policy cannot make full use of the intra-domain links. Suppose that $B$ first requests a block from $A$, say it is $a$. After that, $C$ requests a block from $A$. Since block $a$ already has two copies in $C$'s neighborhood, $C$ will not request block $a$, but can request $b$, $c$, or $d$ from $A$. Let's say it is $b$. Then $D$ tries to request a block from $A$. Since $D$'s neighborhood is $\{A,C,E\}$, $D$ determines that $a,c,d$ are all locally rarest ones, so $D$ may request $a$ from $A$ with probability $1/3$. Similarly, $E$ could request $b$ from $A$ with probability $1/3$. Now, the links between $B$-$C$, $C$-$D$, and $D$-$E$ can be utilized for another round of block exchange. However, after that, $B$, $C$, $D$, $E$ will all own the blocks $a$ and $b$. New data blocks should be transmitted from $A$ through the inter-domain links.

If network coding is used as is shown in Fig. 1(c), the above problem can be largely avoided. With network coding, $A$ will respond each data request with a distinct random coded block. When the finite field is as large as $F(2^8)$ or $F(2^{16})$, the four blocks sent to B-E will be linearly independent with high probability [29], and hence can be used to reconstruct the original blocks. So in subsequent rounds, the data exchange can happen within ISP-2 by locality-aware downloading, resulting in the most efficient use of inter-domain links.

To summarize, the above example shows that data scheduling algorithm can have significant impact on the effect of P2P traffic localization. In this simple example, with random data block scheduling, the probability that a block travels into a domain multiple times is very high. Even with local rarest first data request policy, this probability is not negligible. But if data is scheduled by network coding, this probability will be very low.

## IV. MODEL

In this section, we describe our model for the P2P content distribution system.

### A. Model of the P2P content distribution network

In a peer-to-peer cooperative system, since the server holding the source file does not have the capacity to serve all the users simultaneously, it splits a file, or part of a file(in the case of P2P streaming), into $n$ original blocks $X = (x_1, x_2, ..., x_n)$ and uploads blocks to different clients. The blocks can then be exchanged between the users so that the stress on the server can be substantially alleviated. In order to reconstruct the original file, the users form an application layer overlay cooperative network to collaborate with each other to assemble all the $n$ blocks.

In a large-scale content distribution system such as BitTorrent, each user is assumed to only know the identifies of a small subset of all the participants, which are defined as its neighborhood. The neighboring relationship is assumed to be symmetric, i.e., if node $A$ is in the neighborhood of $B$, then also, $B$ is in the neighborhood of $A$. Each node can exchange metadata information only with its neighbors. The size of the neighborhood is normally a small value.

The number of blocks a peer can concurrently download and upload is constrained by its download and upload capacity respectively. When a peer's upload capacity is saturated, the peer can no longer accept new data requests until part of its upload capacity is freed.

We assume there is a centralized server that keeps track of all the active users in a download session, similar to the tracker concept. When a node wishes to join the session, it contacts the centralized server at the first place to obtain a subset of other users that are already in the system. The new user then tries to connect to each of them to construct its neighborhood. Since we are primarily focus on the potential performance gain when locality information approaches its limit, we assume a locality-aware overlay network construction in our model. We do not care about whether locality information is obtained by peer's reverse-engineering or ISPs' services as they can be seamlessly integrated into this model. We assume a two layer network, i.e., an overlay network on top of the underlying network. Although different granularity of underlying locality information can be obtained, in this paper we only focus on the coarse-grained autonomous system level information, since it is crucial for the study of inter-domain traffic redundancy. So in this model, an underlying AS graph is first presented, then each peer is assigned with an ASN. The overlay network is then constructed in such a way that peers prefer internal peers over external peers, similar to [14]. This topology is then used for our comparison studies between the network coding based approach and the approach that does not use network coding.

### B. Content propagation

*1) LA+LR:* When network coding is not used, we assume a peer knows the block availability information of its neighbors. In reality, this can be realized by periodical buffer map exchanges between neighboring peers. The buffer map describes a peer's available blocks. When a peer is going to request a block, it bases on the local-rarest first policy to make this decision, i.e., it requests the rarest data block in its neighborhood. When there are multiple neighbors that own the rarest data block, the peer applies a locality-aware downloading policy, i.e., it chooses the closest neighbor to download the block. When a peer receives a data block request, it simply transmits the requested block to the requestor.

*2) Pull-based LANC:* Due to the simplified data scheduling algorithm, the conventional wisdom in network coding based P2P content distribution system is that these systems can utilize the push mode to take full advantage of network coding for throughput improvement. However, as T.Small demonstrated [36], this may result in nontrivial linearly dependent redundant data blocks arriving at the same peer, under-utilizing the link bandwidth. Since reducing the traffic is our topmost objective, we use the receiver-driven pull mode for data propagation.

In our pull-based LANC, each peer periodically exchanges global coding coefficients[1] of its available blocks with its neighbors, instead of the simple buffer map information. So, assume a peer $i$'s coding coefficients for its available blocks at time $t$ is represented as a matrix $A_i(t)$, then for any $l$ in $i$'s neighborhood $B(i)$, $i$ knows $A_l(t)$.

When a peer $i$ wants to request a data block, it first enumerates through its neighbors to construct a candidate list that can provide innovative blocks for itself at present. For each neighbor $j$ whose uploading capacity is not saturated, $i$ first infers whether $j$ has innovative blocks for itself by the outcome of the following equation:

$$r = \left| \begin{matrix} A_i(t) \\ A_l(t) \end{matrix} \right| - |A_i(t)|$$

where $|\ |$ denotes the rank of the matrix.

- If $r > 0$, then $i$ determines that $j$ can have $r$ linearly independent blocks for itself. Suppose currently $q$ blocks are in transmission from $j$ to $i$, then $i$ assumes that it can request at most another $r - q$ blocks from $j$ at present;

---

[1] A coded block $b$'s global coding coefficient $g_b$ is an $n$ tuple $(g_1, g_2, ..., g_n)$ that satisfies $b = g_b X^T$.

- If $r = 0$, $i$ simply skips $j$ and moves to the next neighbor.

If the candidate list is not empty, peer $i$ then bases on the locality-aware downloading policy to select a closest neighbor from the candidate list to request an innovative block. This feature makes LANC different from the network coding in clustered topologies. In [32], the authors showed that network coding is superior than non-coded approaches in a clustered topology. However, in their approach, the locality information is only used in the topology construction phase, manifested by the clustered topology. While in LANC, the locality information is used in both the topology construction and downloading decision processes.

Upon receiving a data block request, a peer independently chooses $m$ blocks $b_1, b_2,..., b_m$, generates $m$ random local encoding coefficients $c_1, c_2,...c_m$, makes a linear combination $b = c_1 \cdot b_1 + c_2 \cdot b_2 + ... + c_m \cdot b_m$ of these blocks, and sends the coded block $b$ to the requestor. In case the peer's available number of blocks is less then $m$, it simply uses all its available blocks to generate a random coded block. The global encoding coefficient $g_b$ of block $b$ will be embedded in the coded block, which can be recursively calculated as follows:

$$g_b = \begin{pmatrix} c_1 & c_2 & ... & c_m \end{pmatrix} \begin{matrix} g_{11} & g_{12} & ... & g_{1n} \\ g_{21} & g_{22} & ... & g_{2n} \\ ... & ... & ... & ... \\ g_{m1} & g_{m2} & ... & g_{mn} \end{matrix}$$

where $(g_{i1}, g_{i2}, ..., g_{in})$ is the embedded global coding coefficient of block $b_i$. $m$ is a parameter called encoding density, which directly relates to the encoding complexity. When $m = 1$, it is equivalent to not using network coding at all.

There are two possible ways for the peer to select the $m$ blocks to generate a coded block. If the peer randomly selects $m$ blocks, the scheme is termed as LANC-random. However, this scheme may cause the upstream peer to generate linearly dependent coded block for the requestor. An alternative way is for the requestor to definitely inform the upstream peer to generate a linearly independent block for itself. Indeed, when the requestor $i$ conducts Gaussian elimination to infer whether the upstream peer $j$ has innovative blocks for itself, $i$ actually knows which blocks on $j$ are innovative for $i$. Therefore, $i$ can inform $j$ to decisively select a non-zero coefficient for the specific block. The Gaussian elimination outcome can be easily used for this purpose. The corresponding block of $j$ whose coding coefficients are not all zero after the elimination process is an innovative block for $i$. This scheme is termed as LANC-informed. As we will show later, only when $m$ is very small do LANC-random and LANC-informed show significant difference. Without otherwise stated, the LANC used in this paper refers the LANC-random scheme.

Surely, the linearly dependent redundant block problem cannot be completely avoided by this means, because different peers may accidently produce linearly dependent blocks, however, the chance is very rare.

Finally, when a peer receives $n$ linearly independent coded blocks, it can decode the original blocks $X$ by $X = A^{-1}Y$, where $Y$ is the vector of $n$ received coded blocks, and $A^{-1}$ is the inversion of the matrix induced by the global encoding coefficients of $Y$.

### C. incentive mechanism

An important problem of P2P system is that many users (often called free-riders) are leeching the system, but not contributing resources to the system. In order to discourage free riders, many P2P networks have introduced incentive mechanisms. We consider in this paper a Tit-for-Tat incentive mechanism inspired by the Bittorrent to discourage free-riders. A peer $i$ will not accept a neighbor $j$'s data request when the number of blocks $i$ has uploaded to $j$ minus the number of blocks $i$ has received from $j$ exceeds a threshold value $C$. Only when the upload/download imbalance returns back to within $C$ blocks can $i$ reconsider the data request from $j$.

Heuristically, network coding based approach can increase the diversity of data blocks between neighbors so that tit-for-tat will not choke as much links as in the approach that does not use network coding.

## V. EXPERIMENTAL EVALUATION

In this section, we study the performance of a P2P content distribution network based on LANC and compare it with the state-of-art technique. The approach used for comparison represents the most effective way for P2P traffic localization with mesh-based topology. This approach also utilizes locality-aware overlay network topology construction and downloading, and makes use of the local rarest first policy for data scheduling. We denote this scheme as LA+LR. As demonstrated by [14], this approach can effectively localize traffic in BitTorrent-like systems. We don't consider the random data scheduling policy since it is demonstrated to have much poorer performance than LA+LR [14]. In section V-D, we will also compare LANC with another network coding based approach that only uses locality information in the topology construction phase. The following metrics are used for performance evaluation purpose:

1) inter-domain traffic redundancy(IDTR): measured by the expected number of times each block travels into a domain. Ideally, only one copy of the original file has to travel into a domain and all the peers within this domain can then assemble and reconstruct the original file through cooperation. This ideal case corresponds to the optimal inter-domain traffic redundancy 1, which is very hard, if not impossible, to achieve with mesh-based topology. This optimal traffic burden can be realized if cross-domain IP-layer multicast is supported, or if an application-layer multicast tree is elaborately built such that for each edge in the spanning tree of the underlying AS graph, there is only one overlay link between the two ASes. This tree topology, however, is hard to maintain. In addition, a single-tree based approach will take much longer time to distribute the blocks to all users as it cannot fully utilize the vacant network resources.

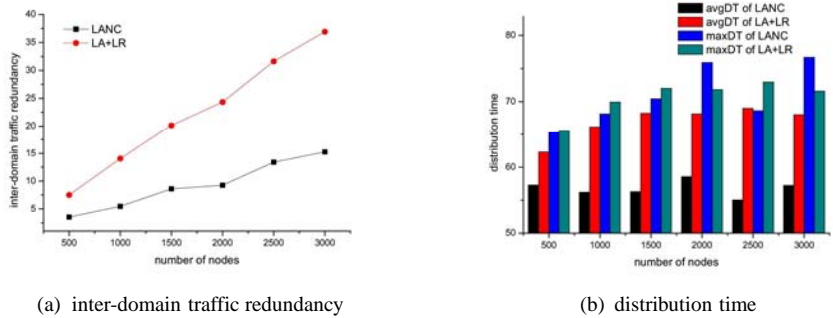

(a) inter-domain traffic redundancy    (b) distribution time

Fig. 2: Impacts of population size

2) average distribution time(avgDT): measured by the average time for the whole population to finish downloading the original file. If there are peers unable to finish the downloads, e.g., in the dynamic setting, these peers are excluded from the calculation of expected distribution time.
3) maximal distribution time(maxDT): measured by the maximal time for a peer to finish the download. Again, this calculation does not consider peers that are unable to finish downloading.
4) number of peers unable to finish the downloads.

The first metric directly relates to ISPs' concern, i.e., how efficient network resources are utilized, and the latter three are indicative metrics for application-level performance.

To study the performance of potentially large number of users under various settings and scenarios, we have implemented a simulator of cooperative system that uses both LANC and LA+LR for content distribution. Our purpose is not to construct a perfectly realistic simulation, but to demonstrate the advantages of locality-aware network coding in P2P traffic localization, which we believe, are of practical importance.

The input to the simulator includes: (1) a locality-aware overlay topology on top of an underlying autonomous system level topology; (2) the upload and download capacities of each participating nodes; (3) the server holding the original file; (4) the size of the file to be distributed. The same input is used for both the simulations of LANC and LA+LR. For each input, the simulation is run for multiple times and the results presented in this paper are the average of these simulations.

The underlying network is the 4-core subgraph[2] of the Chinese AS-level topology measured by our previous effort [37], which contains 37 ASes and 156 edges. The average shortest path length of the underlying AS graph is 1.91. We assign an ASN to each peer. The number of peers in an AS varies according to the degree of the AS. In our experiment, the number of nodes in an AS is set to be proportional to the degree of this AS to reflect the fact that larger ASes often have more Internet users. The peers are connected in such a way that the expected percentage of the intra-domain overlay links

[2] k-core decomposition is believed to be a good approach for identifying the core of a network, see [39] for details.

is fixed to be $p$. The capacities are measured by the number of blocks that can be downloaded/uploaded simultaneously. Link delay is assumed to be uniformly distributed in [0.75ts, 1.25ts], where $ts$ is a time unit.

The simulator is event-driven. When a time unit elapses or when a block downloading finishes, the peer begins a new attempt to request new data from its neighbors. If a peer chooses a new block to download, it models the downloading process by posing a new *receiving* event to be handled in the near future. When all the peers are unable to request new blocks and the event queue is empty, the simulation stops.

To simulate a tit-for-tat scenario, the simulator keeps track of the difference between uploaded blocks minus downloaded blocks from a source $S$ and a destination $D$. If the difference is larger than the pre-configured value $C$, then $S$ will not send any block to $D$ even if there is spare upload capacity at $S$ and spare download capacity at $D$ until $D$ has uploaded enough blocks to $S$.

For each simulation, we record the amount of inter-domain traffic as well as the finish time of each peer. When a peer downloads a block from a neighbor located in another autonomous system, the amount of inter-domain traffic generated equals to the number of hops of the shortest path in the underlying AS graph. For simple reasons, we do not consider the commercial inter-domain routing policy issue here, i.e., whether the path should be valley-free or not [38]. The presence of routing policy will not affect our main result. When the simulation stops. we count the number of peers that cannot finish the downloads. The inter-domain traffic redundancy is then calculated by dividing the amount of inter-domain traffic by the minimum amount of inter-domain traffic.

*A. default setting*

In this subsection, we describe our default setting. The default number of peers is 1000, with average degree of 10. The default ratio of intra-domain links is 70%. The file to be distributed contains 100 blocks, so the minimum amount of inter-domain traffic is $100 \times 36 = 3600$ blocks. Incentive mechanism is disabled in the default setting. The encoding density is *all*, which means to encode all available blocks of a peer. Peers have homogenous capacities, i.e., to be able to upload/download 3 blocks simultaneously. In the default setting, all nodes are preexisting in the network to begin a download session, similar to the flash crowd scenario [14]. Peers will stay in the network after the downloads. Scenarios for network dynamics will be detailed in sectionV-H. The original file server is located in AS17964.

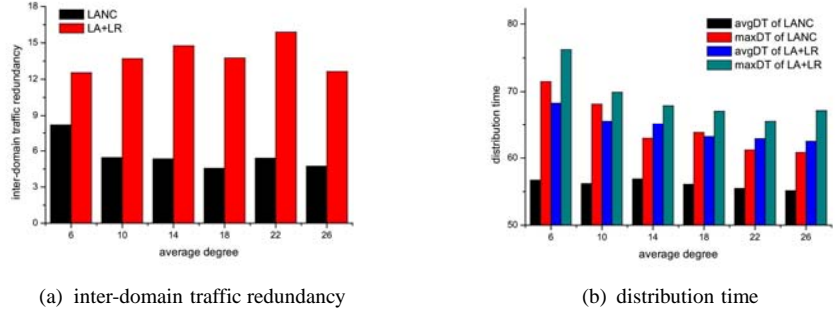

(a) inter-domain traffic redundancy

(b) distribution time

Fig. 4: Impacts of average number of neighbors

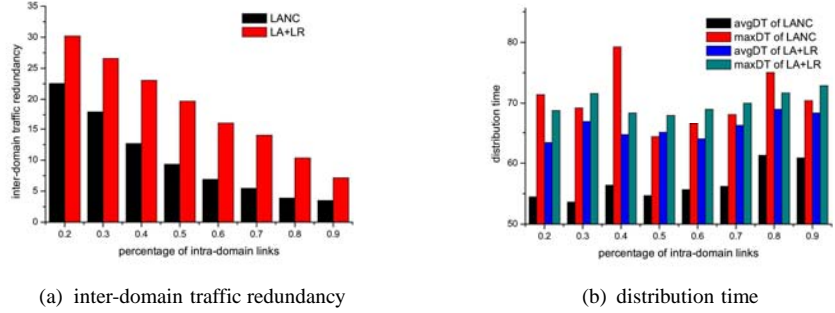

(a) inter-domain traffic redundancy

(b) distribution time

Fig. 5: Impacts of locality awareness

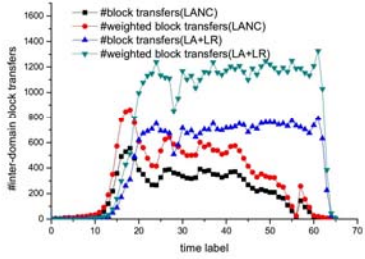

Fig. 3: Temporal occurrence of the inter-domain traffic

Without otherwise stated, we use the default setting for all aspects. In the following, we will investigate how different ingredients affect the performance metrics.

*B. basic performance gain*

Fig. 2(a) reports the inter-domain traffic redundancy and distribution time for different number of nodes. We increase the number of participating nodes from 500 to 3000. Fig. 2(a) shows that LANC can effectively reduce the inter-domain traffic redundancy. In most cases, compared with LA+LR, LANC can reduce the inter-domain traffic redundancy by over 50%[3]. As the number of users increases, the inter-domain traffic redundancy increases accordingly for both schemes, but the growth trend of LANC is slower than LA+LR. Fig. 2(b) shows that the average distribution time of LANC is smaller than LA+LR, indicating better user perceived application-layer performance, which is also confirmed by [32].

Fig. 3 reports the amount of inter-domain traffic transfers occurred in different time slots. In addition to the number of transfers, we also report the weighted number of transfers, that is, each transfer is multiplied by its number of inter-domain hops. It is shown that in LA+LR, the inter-domain traffic remains stable after an initial warm-up stage. But in LANC, after the initial warmup period, the amount of inter-domain traffic reaches its peak, then it starts to decrease gradually. This is because after the initial stage, the probability for a peer to find innovative blocks in its same domain increases quickly.

*C. density of overlay network*

Fig. 4 reports how the overlay network density affects the inter-domain traffic redundancy and distribution time. We model the overlay network density by tuning the average number of neighbors a peer can have, which is increased from 6 to 26. It is shown that generally, the inter-domain traffic decreases as the average number of neighbors increases for LANC. However, for LA+LR, increasing the overlay network density has no obvious effect on traffic localization improvement. When the average degree is larger than 10, the inter-domain traffic redundancy of LA+LR typically is three times of LANC. Fig .4(b) also shows that the average distribution time slightly decreases as the overlay network becomes denser. This superiority of LANC also comes from the randomization introduced by network coding. With LANC, when the network gets denser, the opportunity for a peer to get innovative blocks from within its domain increases slightly. While with LA+LR, although the denser topology can increase the number of neighbors that have useful blocks for a peer, but often the locally rarest one resides outside its domain, thus enforcing it to retrieve the block from non-local neighbors.

---

[3] The reason why the absolute inter-domain traffic redundancy is low(slightly above 3) in [14] is that it uses a very stringent locality-aware policy, i.e., it selects only $k=1$ external neighbor of the total 35 neighbors. In the case when $k=5$, the traffic redundancy reaches around 10, which is comparable with our results of LA+LR.

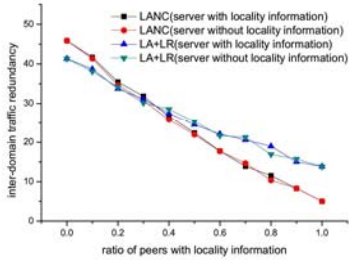

Fig. 6: Impacts of incremental locality service deployment

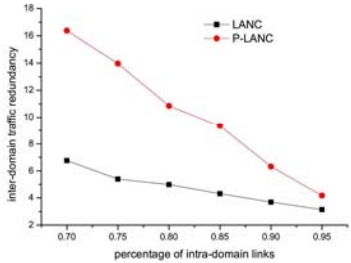

Fig. 7: Impacts of locality information usage redundancy

Nevertheless, though larger number of neighbors could reduce the inter-domain traffic redundancy for LANC, it also means more state information to be maintained for neighbors, more control message overhead between neighboring peers and more complex data scheduling process. Fig .4 shows that the performance will not increase much when the average degree reaches a certain value, e.g., 10. This means, in reality, a peer does not have to maintain very large number of neighbors to achieve fairly good performance.

### D. locality awareness

Improving the accuracy and granularity of locality information centers many previous P2P traffic localization efforts. In this subsection, we show how locality information can affect different metrics. We simulate three scenarios. The first is to simulate the increasing accuracy of locality information, the second is to simulate the incremental deployment of the locality information, and the third is to simulate different usage of locality information.

*1) accuracy of locality information:* The accuracy of locality information is modeled by the ratio of intra-domain links among all the overlay network links. Intuitively, the larger this ratio becomes, the more accurate the locality information is. Fig. 5(a) shows the inter-domain traffic redundancy for LANC and LA+LR, with the intra-domain link percentage ranging from 20% to 90%. It is evident that as the accuracy of locality information increases, the inter-domain traffic redundancy decreases for both schemes. Fig. 5(b) reports the distribution time for both schemes, showing that the average as well as the maximal downloading time are largely unaffected as the accuracy of locality information varies from one extreme to the other, except for a slight increase when $p > 70\%$. So in reality, whenever possible, LANC can benefit from more accurate locality information without sacrificing application-layer performance.

*2) incremental deployment of locality service:* We model the incremental locality service deployment by varying the ratio of peers with locality information. For locality-aware peers, their neighbors are chosen as previously described with the probability $p$ of intra-domain neighbors to be 70%, and for peers without locality information, their neighbors are randomly chosen. Fig. 6 shows how the inter-domain traffic redundancy is affected by the incremental deployment of locality service. For each scheme, we simulate two cases: the locality service is available/unavaiable to the server. For both the LANC and LA+LR, the inter-domain traffic redundancy decreases as the locality service is incrementally deployed. It is shown that LANC can benefit more from the aggressive deployment of locality service. This is due to the fact that network coding's randomization enhances the probability for a peer to find innovative blocks from its proximate neighbors, while the locally rarest first policy slows down a peer's increasing probability to fetch useful blocks from its proximity.

*3) locality information usage:* Finally, we compare between two locality usage scenarios with network coding: locality information is only used in topology construction and locality information is used in both topology construction and downloading decision. The former case indeed corresponds to using network coding in clustered topology [32], which we simply call P-LANC, meaning partially locality-aware network coding because the locality information is only partially leveraged, while the latter is LANC. In this experiment, we uniformly distribute the peers to all ASes regardless of their sizes. We range the percentage of intra-domain links from 0.7 to 0.95. Fig. 7 shows the inter-domain traffic redundancy for these two locality information usage cases. It shows that by leveraging locality information in both the topology construction and downloading decision phases, we can maximize the benefits of network coding in terms of inter-domain traffic reduction. Although network coding enables randomization of blocks, it doesn't specify how a peer chooses from its candidate neighbors that have innovative blocks for it. In P-

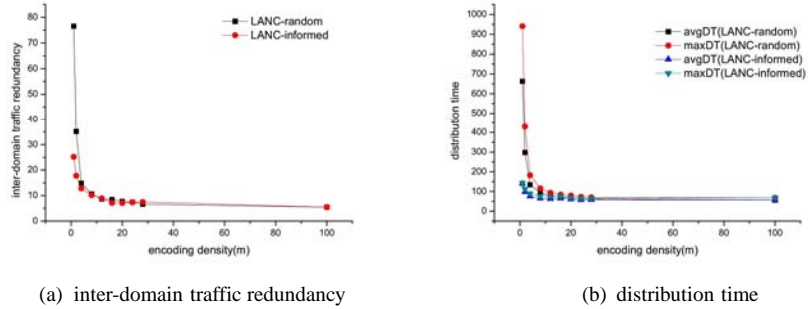

(a) inter-domain traffic redundancy  (b) distribution time

Fig. 8: Impacts of encoding density

LANC, a peer treats the candidates equally, while in LANC, it prefers those neighbors within the same cluster, hence incurring significantly lower inter-domain traffic.

*E. encoding density*

Encoding density is measured by the number of blocks that are used to generate a coded block. Encoding density directly relates to the end user's encoding overhead. Smaller values of $m$ mean lower encoding overhead, however, it also reduces the probability to find innovative blocks from its neighborhood, hence less effective in traffic localization. Actually, when $m$ is too small, by randomly selecting $m$ blocks, a peer may have very high chance to produce a linearly dependent block for the requestor, although it does have innovative blocks for that requestor according to the coding coefficients. So when $m$ is too small, LANC could be outperformed by LA+LR.

Fig .8(a) shows how the encoding density affects the inter-domain traffic redundancy. When $m$ is very small, the traffic redundancy is very high. However, inter-domain traffic redundancy decreases sharply as $m$ increases from 1 to 4. In reality, $m$ has not to be very large to achieve nearly optimal performance. When $m \geq 20$, the performance gain is nearly the same as encoding all blocks.

Fig.8(b) shows the average and maximal distribution time of all peers. These two curves have similar trends as the inter-domain traffic redundancy. When $m$ is very small, the distribution time could be an order of magnitude longer than that of median $m$. However, it decreases sharply as $m$ increases. The prolonged distribution time is largely because of the linearly dependent blocks generated with smaller $m$, wasting network resources.

One way to reduce the linearly dependent blocks is to use the previously mentioned LANC-informed scheme, i.e., let the requestor inform the sender which block is innovative for itself so that the sender can choose a non-zero coding coefficient for that specific block. This scheme can indeed lower the inter-domain traffic redundancy for small $m$, as is shown in Fig .8(a). However, the inter-domain traffic redundancy is still higher for small values of $m$. This is because although the upstream peer will now definitely produce useful blocks for its direct downstream peers, small values of $m$ reduces the potential usability of the coded block for its farther offsprings, e.g., peers two hops away. On the other hand, when $m$ is large, the coded block will not only have local utility, but will also have far-reaching global utility.

When $m$ reaches a certain value, e.g., when $m \geq 8$, the difference between LANC-random and LANC-informed becomes indiscernible. This means in reality, LANC-random with median encoding density is better than LANC-informed as LANC-random is easy to implement and eliminates the need to exchange the informing messages.

*F. heterogeneous capacities*

In real world, node can often have different upload and download capacities. Most of the nodes will be DSL users with relatively lower capacities, while some nodes will have much higher capacities, e.g., university users with broadband Ethernet access. In this experiment, we randomly assign 10% nodes with higher node capacities. These nodes are 10 times faster than other nodes, which means, they can upload/download 30 blocks concurrently. We study two cases: the server is assigned with low capacity and is assigned with high capacity respectively.

Table I shows how network heterogeneity affects the system performance for both LANC and LA+LR. It is shown that the network heterogeneity does not affect our main result, i.e., LANC can reduce the inter-domain traffic by 50% compared with LA+LR. In addition, it is observed that if the sever has high capacity, the inter-domain traffic redundancy and the distribution time both decrease significantly, whereas if the server has low capacity, the performance is only slightly improved. This is because with high capacity server, the time for the server to disseminate a copy of the original file drastically decreases. This effect has practical implication for real-world deployment, i.e., to place the original file on a high capacity server whenever possible.

*G. tit-for-tat*

In this experiment, we enable the tit-for-tat incentive mechanism. Fig.9 reports how the threshold value $C$ affects the system's performance. Smaller value of $C$ places more stringent constraint for two neighbors to exchange traffic. It is shown that $C$ has higher impact on the inter-domain traffic redundancy for LANC than for LA+LR. As $C$ increases, the inter-domain traffic redundancy of LANC decreases slowly, whereas it is almost unaffected with LA+LR. On the other hand, the distribution time is only slightly affected for LANC, but significantly affected for LA+LR. When $C$ is small, the distribution time for LA+LR exceeds $100ts$, nearly two times that of LANC. And, the number of peers unable to finish the downloads is high for LA+LR with small $C$. This means, when tit-for-tat incentive mechanism is used, LANC behaves much better than LA+LR both in terms of inter-domain traffic redundancy and application-layer performance. This result is because when $C$ is small, nodes without network coding are more easily to end up with blocks that are of little interest to their neighbors, so the longer distribution time and higher probability to be choked.

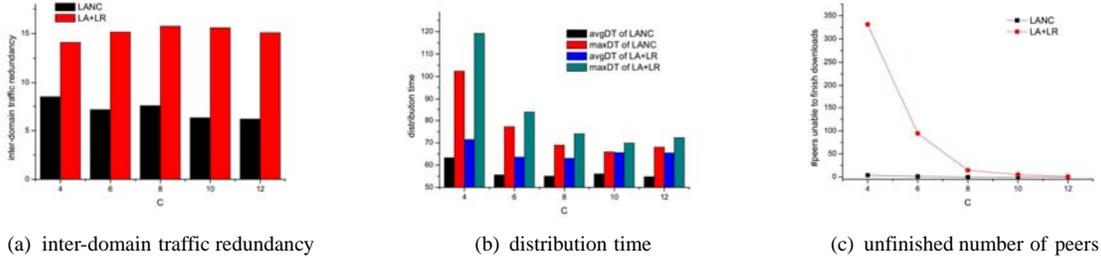

(a) inter-domain traffic redundancy   (b) distribution time   (c) unfinished number of peers

Fig. 9: Impacts of tit-for-tat

TABLE I: Impacts of heterogeneous upload/download capacities

| scheme | | IDTR | avgDT | maxDT |
|---|---|---|---|---|
| LANC | low capacity server | 6.01 | 51.13 | 58.19 |
| | high capacity server | 4.13 | 41.51 | 59.86 |
| | no heterogeneity | 5.45 | 56.18 | 68.15 |
| LA+LR | low capacity server | 13.98 | 51.13 | 62.35 |
| | high capacity server | 8.04 | 41.51 | 54.66 |
| | no heterogeneity | 14.05 | 66.18 | 69.95 |

TABLE II: Performance impacts of network dynamics for both LANC and LA+LR.

| Dynamics Category | scheme | IDTR | avgDT | maxDT | UNFN% |
|---|---|---|---|---|---|
| scenario A | LANC | 7.85 | 82.62 | 145.26 | 11.3% |
| | LA+LR | 12.41 | 87.43 | 141.07 | 11.8% |
| scenario B | LANC | 7.44 | 81.85 | 140.82 | 12.8% |
| | LA+LR | 12.69 | 87.80 | 143.83 | 10% |
| scenario C | LANC | 7.91 | 81.63 | 140.14 | 11.8% |
| | LA+LR | 12.04 | 46.17 | 58.45 | 82.3% |
| scenario D | LANC | 4.58 | 64.50 | 85.10 | 29.9% |
| | LA+LR | 10.14 | 70.28 | 75.19 | 32.9% |

## H. network dynamics

Previous experiments all assume a static topology, in which all peers join the network at once to begin a downloading session. These experiments demonstrate LANC's ability for P2P traffic localization. Whereas in reality, P2P networks display high dynamism. Peers join and leave the network with high frequency. In this experiment, we study how LANC and LA+LR perform under various dynamic network environment.

We consider four dynamic scenarios, referred as scenario A, scenario B, scenario C and scenario D respectively. In scenario A, every $10ts$, 100 peers join the network in a batch. Each peer departs the network $10ts$ after it finishes the download. The server is always available. Scenario B only differs one aspect from scenario A, that is, the server leaves the system $10ts$ after all the nodes arrives. In scenario C, all other setting are also the same as scenario A, except that the server takes extreme behavior, i.e., the server departs the system after serving 120 blocks. In these three scenarios, when a node wants to join the network, it creates its neighborhood with the nodes that are already in the network. It follows the locality-aware overlay network building philosophy, such that the ratio between its intra-domain neighbors and inter-domain neighbors is 7:3. In scenario D, we assume that initially all nodes preexist in the network, and in the following, 50 random nodes leave the network every $10ts$, even these nodes do not finish the downloads. This process continues until $100ts$, resulting in a total departure of 500 nodes.

Table II shows how the two schemes perform under these dynamic scenarios. It is shown that network dynamics do not affect our main results, i.e., LANC can achieve much better P2P traffic localization than LA+LR. In the extreme case of scenario C, we observe that with LA+LR, only 17.7% of the peers can finish the downloads, while with LANC, nearly 90% nodes can still finish the downloads. This means from the application-layer perspective, LANC can adapt to sudden server departure much better than LA+LR.

## VI. IMPLEMENTATION CONSIDERATIONS

In this section, we propose some implementation considerations that are of relevance to implement a real LANC based P2P content distribution system.

### A. architectural overview of a LANC peer

Fig. 10 presents an architectural overview for a possible implementation of a LANC peer. The encoder, decoder and the coded blocks buffer are the same as a typical network coding based P2P content distribution system. Coded blocks are assembled and stored in the peer's buffer. When the peer is to generate a coded block, the encoder randomly selects $m$ blocks from the coded block buffer and generates a coded block. After assembling $n$ linearly independent coded blocks, the decoder can begin the decoding process.

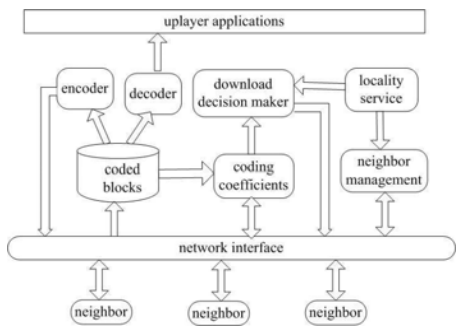

Fig. 10: An architectural overview of a LANC peer implementation.

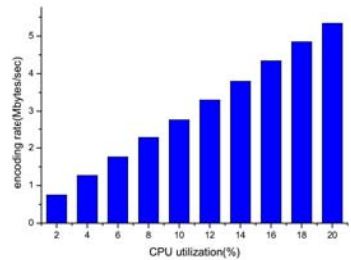

Fig. 11: Encoding rate under different CPU utilization.

The function of neighbor management is to switch a peer's neighbors when the existing neighbors cannot satisfy its downloading requirement. A rule of thumb in neighbor switching is to maintain the locality awareness through the locality service. Neighbor switching, however, is not considered in our simulation because it is not our main focus. The coding coefficients is similar to the traditional buffermap in mesh-based P2P systems, used for exchanging the block availability information between the neighboring peers. The download decision maker is at the heart of LANC, which relies on both the coding coefficients and locality service to make the download decision. The locality service will be detailed below.

### B. integrating locality service

The locality service is a key aspect for the success of LANC as it is used by the initial overlay topology construction, neighbor management and download decision maker. In the previous parts, we focused only on the autonomous system level topology as this information is easy to obtain and can even be embedded in the client software. However, other kinds of locality information or locality service can be seamlessly integrated into this implementation to better utilize the network resources. The peer can rely on active probing, prefixing matching, network coordinate system, or ISP-offered services such as P4P and Oracle, to get more fine-grained locality information. More fine-grained locality information enables more fined grained granularity of traffic localization, e.g., localizing the traffic within the same university or same city.

The locality-aware topology construction can be implemented by the tracker [1], [2], [14], [20], the ISP's proxy tracker near the network boundaries [9], [14], [19], or by the peer itself. The tracker maintains all the active participating nodes in a download session. Upon receiving a joining request or a neighbor switching request, the tracker can allocate a list of peers that are close to the newcomer. The ISP's proxy tracker can intercept a tracker's response to a peer's neighbor request, and modify the response neighbor list with the peers that are considered to be close to the requestor by the proxy tracker.

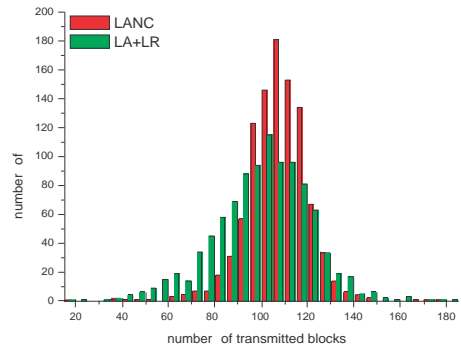

Fig. 12: Distribution of the number of blocks transmitted during a download session.

### C. encoding overhead

The encoding and decoding overhead is one of the major concern for users. In Avalance [35], the average CPU usage is reported to be 20% for a typical downloading session. In [25] and [34], the CPU is tuned to 100% to study the maximal encoding and decoding rate.

In this paper, we study the computation overhead from another angle. We present a theoretical analysis of per byte encoding overhead in terms of the number of multiplication and addition operations taken place over the Galois Field, which offers us a means to evaluate how different aspects and settings affect the computation overhead. In addition, we present a practical evaluation of the achievable encoding rate under certain CPU utilization, e.g., under 20%, which we believe is reasonable for users's adoption.

Assume a file of size $Q$ is split into $n$ blocks, each of size $k$. Generating a coded block consists of three computation intensive operations: (1) generate $m$ random local coefficients, (2) encode $m$ blocks with the $m$ local coefficients, and (3) compute the global coding coefficients. The coefficients generation is often negligible. Encoding $m$ blocks incurs $mk$ multiplication operations and $(m-1)k$ addition operations. Computing the global coding coefficients incurs another $mn$ multiplication operations and $(m-1)n$ addition operations. As a result, generating a coded block of length $k$ requires $m(k+n)$ multiplication operations and $(m-1)(k+n)$ addition operations in total, which is equivalent to $m(1+\frac{n}{k})$ multiplications and $(m-1)(1+\frac{n}{k})$ additions per byte. When the block size $k$ is not too small, then typically $n ¿ k$, for instance, when $k$=128k, then a 1G file will result in $n$=8k. In this case, the computation overhead is dominated by $m$. However, if $k$ is too small, then $n/k$ will play an important role in the computation overhead.

We provided an optimized implementation of the encoding process and evaluated the encoding rate under certain CPU utilization. Our implementation is on the Galois Field $F(2^8)$. We used a table of 64K(256×256) to store the multiplication result. Later on, when we encounter a multiplication, we can index the table instead of really doing the multiplication. This improves the encoding rate since our experience shows that multiplication is much slower than table indexing. We don't use the table indexing for the addition operation since XOR operation is slightly faster than the table indexing. Fig. 11 presents our test result for the achievable encoding rate with CPU utilization under 20%. In our experiment, the block size is 64K, and the number of blocks is 1600. The encoding density $m$ is 20. We used a typical PC with CPU rate 1.86G. The result is quite impressive. Even when the CPU utilization is 4%, the encoding rate can achieve 1.2Mbytes/sec.

Fig. 12 presents the distribution of the number of blocks transmitted by each peer during a typical downloading session. It is shown that the number of blocks transmitted by each peer in LANC is more centralized around the expected value than LA+LR, indicating a more balanced uploading traffic load on each peer, hence similar encoding overhead for each node. However, we also noticed that the overall number of transmitted blocks by LANC is about 10% more than that of LA+LR. The additional transmissions are all linearly dependent blocks. This issue will be addressed in our future work.

Discussion of decoding overhead is omitted here since it is reported to incur similar or lighter overhead compared with the encoding overhead [34].

*D. ISP caching compatibility*

Some may argue that networking coding disables ISP's ability to do caching. This is not true, ISPs can modify their caches with little effort to support network coding enabled P2P systems. Instead of serving the requestor with original hit blocks, the cache now will serve the requestor with encoded blocks. The cache will perform a random coding of some of its cached blocks and send the coded block to the requestor.

Compared with caching the original data blocks, network coding has the potential of releasing the legal liability of ISPs. If the caches are configured in the way that they can only cache less than $n$ blocks of a file, then most original blocks cannot be recovered.

VII. CONCLUSIONS

In this paper, we answered one question: can we do any better in P2P traffic localization beyond P4P, i.e., when the accuracy of locality information approaches its limit? We proposed LANC–locality aware network coding, as an effective approach to relieve the tension between ISPs and P2P content providers beyond P4P. We found using locality information in overlay network construction alone cannot fully leverage the benefits of network coding for P2P traffic localization. By integrating network coding and using locality information in both the overlay network construction and downloading decision phases, LANC can reduce the inter-domain traffic redundancy by over 50% compared with the state-of-art techniques. We carried out extensive simulations and demonstrated that this property holds for different network settings and under various network dynamic environment. The simulation results also provide instructive information for choosing reasonable parameters, e.g., average number of neighbors to be maintained, encoding density, and server placement, for a real LANC implementation. Also, practical implementation issues are considered, including the prototype peer architecture, integration of the locality service, and a fast encoding implementation.

While previous endeavors for wide application of network coding to P2P content distribution are impeded by the concerns about encoding and decoding overhead, we believe that in addition to the application-layer throughput improvement, network coding's powerful ability for P2P traffic localization is worthy the cost of computation overhead. And the encoding overhead is not as high as previously perceived by our fast encoding implementation. It is no doubt that ISPs will welcome LANC due to the significant reduction of traffic redundancy. LANC is also desirable for P2P content providers due to its ability to improve application-layer throughput as well as its ability to relieve the tension between P2P content providers and ISPs. The end users, by contributing some of its computation resources, could also benefit from LANC with lower cost for its traffic, since otherwise, increasing the end users' fee is the only way for ISPs to balance its revenue and investment.

ACKNOWLEDGEMENTS

This work is partly supported by the National Natural Science Foundation of China under Grant No. 60673168 and the Hi-Tech Research and Development Program of China under Grant No. 2008AA01Z203.